%% file: conference_041818.tex
\documentclass[conference]{IEEEtran}
\IEEEoverridecommandlockouts
\IEEEpubid{\makebox[\columnwidth]{978-1-7281-0407-2/19/\$31.00 \copyright 2019 IEEE \hfill} \hspace{\columnsep}\makebox[\columnwidth]{ }}
\input{packages.tex}
\usepackage{subcaption}

\newcommand{\OpenDSS}{OpenDSS\ }

\begin{document}

\title{A Framework for Generating Synthetic Distribution Feeders using OpenStreetMap
\thanks{This work was partially supported by the Office of Naval Research under Award Number N00014-18-1-2393 and the Engineering Research Center Program of the National Science Foundation and the Office of Energy Efficiency and Renewable Energy of the Department of Energy under NSF Cooperative Agreement Number EEC-1041895.}
}

\author{\IEEEauthorblockN{Shammya Shananda Saha\IEEEauthorrefmark{1}, Eran Schweitzer, Anna Scaglione}
\IEEEauthorblockA{\textit{School of Electrical, Computer, and Energy Engineering} \\
\textit{Arizona State University}\\
Tempe, USA \\
\IEEEauthorrefmark{1}shammya.saha@asu.edu}
\and
\IEEEauthorblockN{Nathan G. Johnson}
\IEEEauthorblockA{\textit{The Polytechnic School} \\
\textit{Arizona State University}\\
Mesa, USA 
}
%
}

\maketitle

\begin{abstract}
This work proposes a framework to generate synthetic distribution feeders mapped to real geo-spatial topologies using available OpenStreetMap data. The synthetic power networks can facilitate power systems research and development by providing thousands of realistic use cases. The location of substations is taken from recent efforts to develop synthetic transmission test cases, with underlying real and reactive power in the distribution network assigned using population information gathered from United States 2010 Census block data. The methods illustrate how to create individual synthetic distribution feeders, and groups of feeders across entire ZIP Code, with minimal  input  data for any location in the United States. The framework also has the capability to output data in \OpenDSS format to allow further simulation and analysis.
\end{abstract}



\section{Introduction}
Confidentiality and security policies governing critical infrastructure data can limit researchers from accessing real power systems data needed for scientific development. For decades, researchers have extensively used a small set of standardized networks such as the IEEE transmission and distribution test cases \cite{testfeeder1,testfeeder2}. Recent research has developed additional test cases using algorithms that generate synthetic networks to mimic realistic power grids \cite{grid_structural,activs1,activs2,ABEYSINGHE20172595,eran_synfeeder,RNM, smartds}. These open source works have provided more example networks for scientific exploration. 

The development of synthetic transmission test cases (\emph{ACTIVS}) presented in \cite{grid_structural} begins with publicly available real energy and population data, from which synthetic substations are placed geographically. Buses are subsequently added to these substations, connected by transformers and a network of transmission lines, using an iterative process that considers multiple factors. Statistical validation of these models using typical topological criteria is provided in \cite{grid_structural,activs5}.

\subsection{Prior art on generating synthetic distribution systems}
A recent study developing synthetic radial distribution feeders \cite{eran_synfeeder} views the distribution system as a random graph with nodes and edges that can be imbued with various properties following realistic statistics for the topology and parameters, derived from a large set of real feeders. The approach includes an \textit{analysis} and a \textit{synthesis} step. The analysis step identifies statistical distributions between properties such as load, node, degree or cable length using data from the Netherlands. The statistics are then input to a synthesis algorithm that creates feeders with similar topological and physical characteristics; importantly, the statistical trends for different operating conditions of voltage, phase angles power flows are also realistic. Moreover, \cite{eran_synfeeder} provides a method for validating the results, using metrics of statistical similarity such as the Kulback Leibler distance of the trends of the real and synthetic cases.
\IEEEpubidadjcol
Another line of work is based on the Reference Network Model (RNM) \cite{RNM}, a large-scale distribution planning tool that can help regulators to estimate efficient costs in the context of incentives and regulation applied to distribution companies. The synthetic test-cases developed using RNM allows for the simultaneous planning of high-, medium-, and low-voltage networks using simultaneity factors and lays out cables in urban areas, taking street map into consideration. An adaptation of RNM to create RNM-US is shown in \cite{smartds} which provides full-scale, high-quality synthetic distribution system dataset(s) for testing distribution automation algorithms, distributed control approaches, and other emerging distribution technologies.  Broad application of RNM is hindered by the significant amount of data required such as (a) geo-referenced for transmission, substations, and consumer data, (b) load profiles, (c) equipment library for all power system components, (d) technical and economic parameters governing operation, and (e) environmental and topography data.  

\subsection{Contribution}
The contribution of this paper is threefold: 1) substation locations are selected from the \emph{ACTIVS} models \cite{grid_structural}, a set of geo-embedded synthetic models of the US transmission system, to enable joint evaluation of synthetic transmission and distribution systems; 2) following the assumption that power distribution network follows the road network; \emph{radial positive sequence} synthetic distribution feeder topologies are generated using OpenStreetMap (a constantly expanding publicly available data source for road network) \cite{OpenStreetMap} to create a spatially embedded dataset; 3) loads are assigned to nodes in the distribution network using US census population data. The developed framework also has the capability to provide the distribution feeder model in \OpenDSS \cite{opendss} format allowing simulation capabilities like hosting capacity calculation \cite{hosting_capacity}, complex infrastructure network simulation \cite{complex_infrastructure}, and more. 


\begin{figure}
    \centering
    \resizebox {17pc} {!} { \input{Figs/ProcessFlowv2.tex} }
    \caption{Flowchart representing the synthetic distribution feeder generation framework}
    \label{fig:flowchart}
\end{figure}
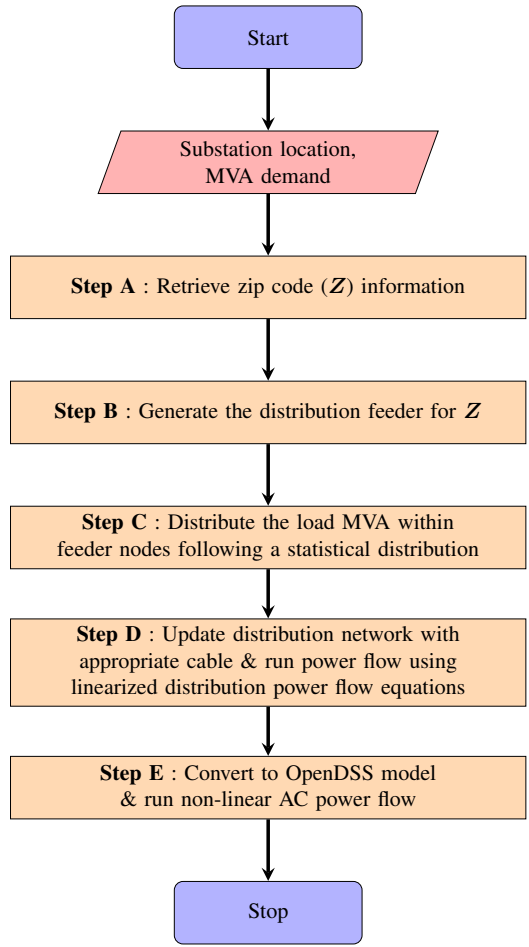


\section{Distribution Feeder Generation for a Single Substation}
The \emph{ACTIVS} cases \cite{grid_structural} are based on a hierarchical clustering algorithm which groups postal codes to $N_s$ number of substations, where $N_s = N_l + N_g + N_b$, the number of substations containing only loads, only generators, and both respectively. Substations with \emph{positive net real power} are used as the source for distribution feeders in this  work.

The set of substations with positive net real power demand is defined as $N_D$; where $N_D \in N_s$. The methods introduced are demonstrated for a substation ($N$) located at $33.3420^{\circ}$ latitude and $-111.6739^{\circ}$  longitude (Mesa, Arizona) with a load demand of $40.47~ + j 11.14~\textnormal{MVA}$. A simplified flowchart of the framework is shown in Fig. \ref{fig:flowchart}. Results from each step are presented alongside methods to illustrate the incremental process for creating synthetic distribution networks. 

\subsection{Finding the ZIP Code ($Z$) of substation $N$}
\begin{figure}
    \centering
    \includegraphics[width=\textwidth]{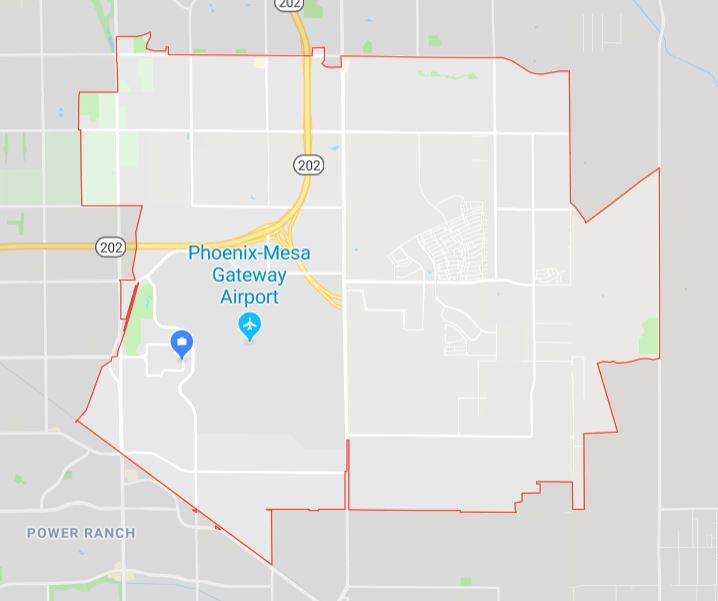}
    \caption{Overview of ZIP Code $85212$ from Google Maps}
    \label{fig:85212}
\end{figure}

Location information (longitude, latitude) of the substation from \emph{ACTIVS} cases is used to identify the corresponding ZIP Code ($Z$) from the "US 2010 Census Bureau ZIP Code" shapefile\footnote{A shapefile is a simple, non-topological format for storing the geometric location and attribute information of geographic features.} \cite{wiki:census_block} or the Google Reverse Geocoding API \cite{reverse_geocode} if unavailable in the shapefile. If both processes fail, then the closest ZIP Code to substation $N$ is assigned. The selected substation for demonstration belongs to the ZIP Code $85212$. Fig. \ref{fig:85212} illustrates the territory of ZIP Code $85212$ from Google Maps.

\begin{figure*}
\begin{tikzpicture}
\node[inner sep=0pt] (subgraph1) at (0,0)
    {\includegraphics[width=.5\textwidth]{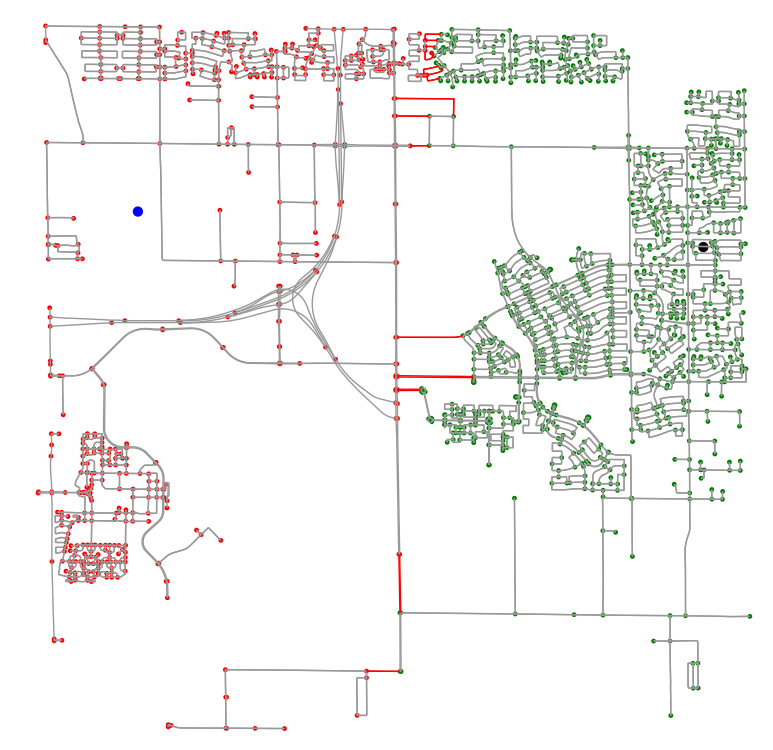}};
\node[inner sep=0pt] (subgraph1edit) at (9.25,-0.2)
    {\includegraphics[width=.5\textwidth]{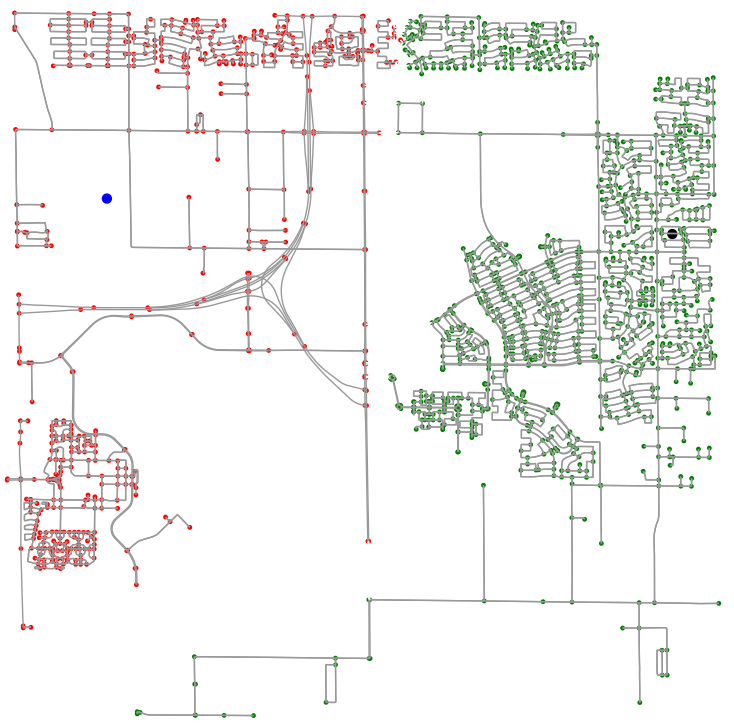}};
\draw[thick]  (-2.75,-4.25) rectangle (0,-3.2); 
\node at (-1.35,-4.55) {Isolated Nodes ($I$)};
\draw[thick]  (6.2,-4.75)  rectangle (9.1,-3.55);
\draw[dashed,thick]  (-4.25,4.25)  rectangle (4.45,-5.25);
\draw[dashed,thick]  (4.75,4.25)  rectangle (13.75,-5.25);
\node[draw= none,align=left] at (0.15,-5.0) {(a)};
\node[draw= none,align=left] at (9.2,-5.0) {(b)};
\end{tikzpicture}
\caption{Creating sub-graphs and connecting isolated nodes to appropriate sub-graph}
\label{fig:isolated_subgraph}
\end{figure*}
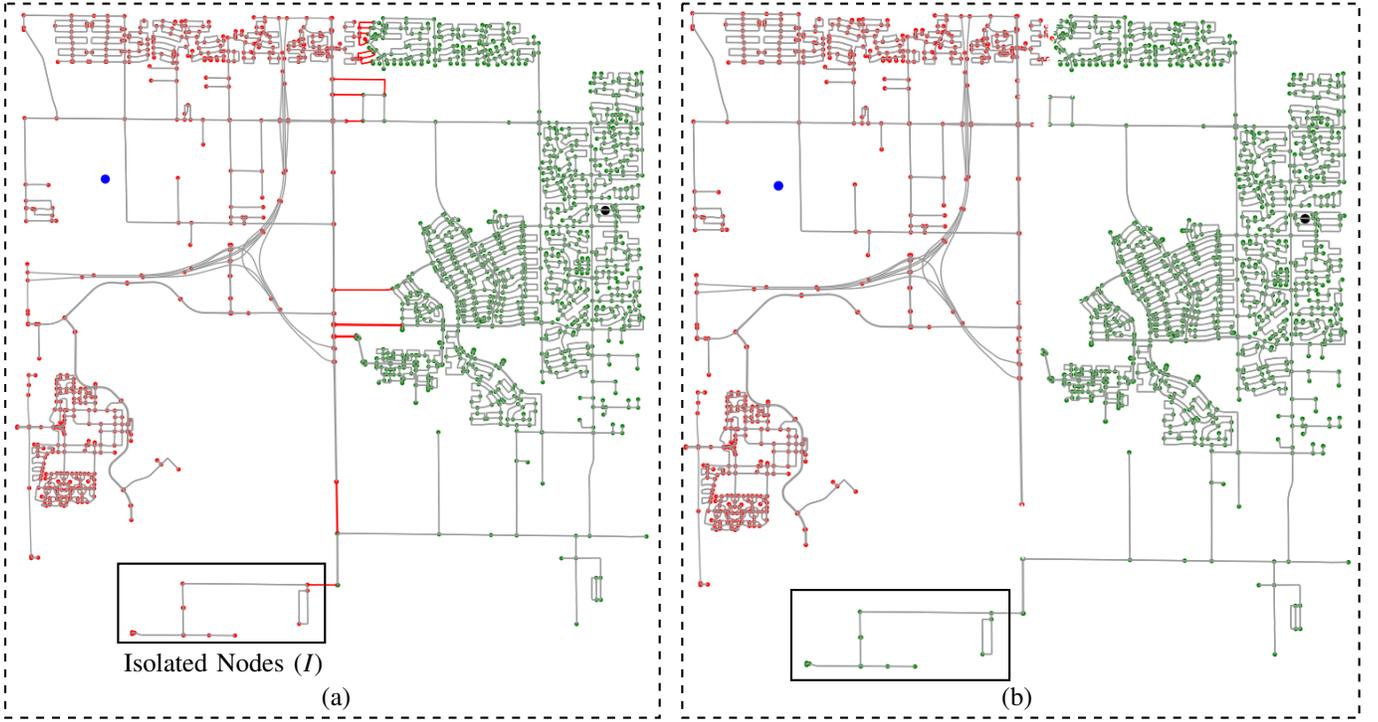


\subsection{Generating the Distribution Feeder Graph for $Z$}
\begin{enumerate}[label=(\roman*)]
    \item A ZIP Code can be served by multiple substations. Using \emph{ACTIVS}, the set of substation(s) ($N_Z$) in $Z$ are retrieved, giving $N \in N_Z \in N_D$. For ZIP Code $85212$, the \emph{ACTIVS} case returns two substations with positive net loads, the second one to be located at $(33.3376^{\circ},-111.5900^{\circ})$ with a load demand of $55.07~ +j 8.51~\textnormal{MVA} $. 
    \item The 'drive' network (drivable public streets excluding service roads) for $Z$ is retrieved using \cite{BOEING2017126}. The retrieved network is a directional graph ($G_Z$) with self-loops and parallel edges.  
    \item $G_Z$ is then split into Voronoi regions \cite{wiki:voronoi_diagram} where the number of Voronoi partitions $\abs{N_Z}$ equals to number of substations in ZIP Code $Z$\footnote{$\abs{\bullet}$ of a set represents the cardinality of the set.}. Fig. \ref{fig:isolated_subgraph}\textcolor{blue}{a} shows $G_Z$ for the ZIP Code $85212$ having two substations with positive net loads. The \textbf{\textcolor{blue}{blue}} and \textbf{black} colored circles represent the location of substations in $85212$. Fig. Fig. \ref{fig:isolated_subgraph}\textcolor{blue}{a} also illustrates the two Voronoi regions corresponding to the two substations. The \textbf{\textcolor{red}{red}} and \textbf{\textcolor{darkgreen}{green}} nodes are distribution feeder nodes inside the \textbf{\textcolor{blue}{blue}} and \textbf{black} substation's Voronoi regions, respectively. The edges marked in \textcolor{red}{red} in  Fig. \ref{fig:isolated_subgraph}\textcolor{blue}{a} are the edges connecting the two Voronoi regions.
    \item $\abs{N_Z}$ number of sub-graphs ($g_z$) are then created from $G_Z$ by splitting along the edges connecting the Voronoi regions. A sub-graph only includes the strongly connected nodes \cite{wiki:scc}; hence the sub-graph creation process can create isolated nodes as shown in Fig. \ref{fig:isolated_subgraph}\textcolor{blue}{a}. The isolated nodes either have degree 0 or are part of an isolated graph. Let $G_Z^n,~ g_z^n$ be the set of nodes for $G_Z$, sub-graph $g_z$ respectively while $I$ is the set of isolated nodes. It follows that $I= G_Z^n \setminus \cup_{z=1}^{\abs{N_Z}} g_z^n $, where $\cup_{z=1}^{\abs{N_Z}} g_z^n $ represents the union of all the sub-graph nodes. Using Algorithm \ref{alg:node_connection} the isolated nodes are connected to the appropriate sub-graph ($g_z$) as shown in Fig. \ref{fig:isolated_subgraph}\textcolor{blue}{b}.
        \begin{algorithm}[h]
        \caption{Algorithm for connecting isolated nodes }\label{alg:node_connection}
            \begin{algorithmic}[1]
            \For {each node $i \in I$}
            \State find $A_i$ where $A_i$ is the set of nodes adjacent to node $i$ in $G_z$
            \State find the $k^{th}$ sub-graph $g_{z,k}$ for the first node $a$ such that $a \in A_i \cap g^n_{z,k}$
            \State connect node $i$ to node $a$ and copy the node and edge properties from $G_Z$ to $g^n_{z,k}$
            \State remove $i$ from $I$
            \EndFor
            \end{algorithmic}
        \end{algorithm}
    \item A minimum spanning tree using Kruskal's algorithm \cite{wiki:Kruskal's_algorithm} is calculated for each $g_z$. 
    \item The census block data for $Z$ is retrieved using US 2010 Census data. If  $B_Z$ is the set of census blocks for ZIP Code $Z$,  there are $\abs{B_Z}$ census blocks within $Z$. There are $662$ census blocks for ZIP code $85212$, i.e. $\abs{B_Z}=662$.
    \item For every node in $\cup_z g_z$, the framework finds the census block that a node belongs to using the latitude and longitude information of the node and geometry of census blocks. 
    \item The framework then assigns a weight ($p_w^n$) to each node based on the population information of the census block the node belongs to. For instance, if node $(n)$ belongs to census block $b_z^{i} \in B_Z$, ~$p_w^n= \dfrac{m(b_z^{i})}{\sum_{i=1}^{\abs{B_Z}}m(b_z^{i})} $, where $m(b_z^{i})$ returns the population of the census block $b_z^{i}$. For $85212$, total population, or $\sum_{i=1}^{\abs{B_Z}}m(b_z^{i})$, is $25015$.
\end{enumerate}

The output of these steps is a set of sub-graphs $\{g_z\}$ with $\abs{N_Z }$ elements where sub-graph $g_{z,N}$ corresponds to substation $N$ and $g_{z,N} \in \{g_z\}$.


\subsection{Assigning load to sub-graph $g_{z,N}$}
\label{assign_load}
\begin{enumerate}[label=(\roman*)]
    \item The node closest to the location of substation $N$ is chosen to be the slack bus/ substation node for distribution feeder for the sub-graph/ feeder $g_{z,N}$. The big \textbf{\textcolor{red}{red}} circle in Fig. \ref{fig:load_no_population} indicates the substation node.
    \item The substation demand is then distributed among the nodes of the sub-graph $g_{z,N}$. Zero load nodes are identified using the methodology in \cite{eran_synfeeder}. The substation node is designated to have zero load. The load distribution for each node under the sub-graph $g_{z,N}$ is equated using the following equations:
    \begin{align}
             P(n) &= P_N \left(\dfrac{1}{n} + \epsilon \right) \times p_w^n  ~;~ Q(n) &= Q_N \left(\dfrac{1}{n} + \epsilon \right) \times p_w^n 
    \end{align}
    Here, $P_N$ and $Q_N$ are the real and reactive power load demand of substation $N$, respectively, and $\epsilon$ can be chosen from any distribution such as uniform distribution, or t-location scale distribution as in \cite{eran_synfeeder}.
    The final $P(n)$ and $Q(n)$ values are then scaled using the following equations to match the total demand $P_N$ and $Q_N$. 
    \begin{align}
            P(n) &= P_N \times \dfrac{P(n)}{\sum P(n)} ~;~ Q(n) &= Q_N \times \dfrac{Q(n)}{\sum Q(n)} 
    \end{align}
    The developed framework does not necessarily require population information for modeling the feeder. Setting the value of $p_w^n $ to 1 for all nodes will ignore the effect of population on the load distribution. Figure \ref{fig:load_no_population} shows the heat-map of real power of $g_{z,N}$ done following a t-location scale distribution while excluding the population information. 
    \begin{figure}
        \centering
        \includegraphics[width=1\textwidth]{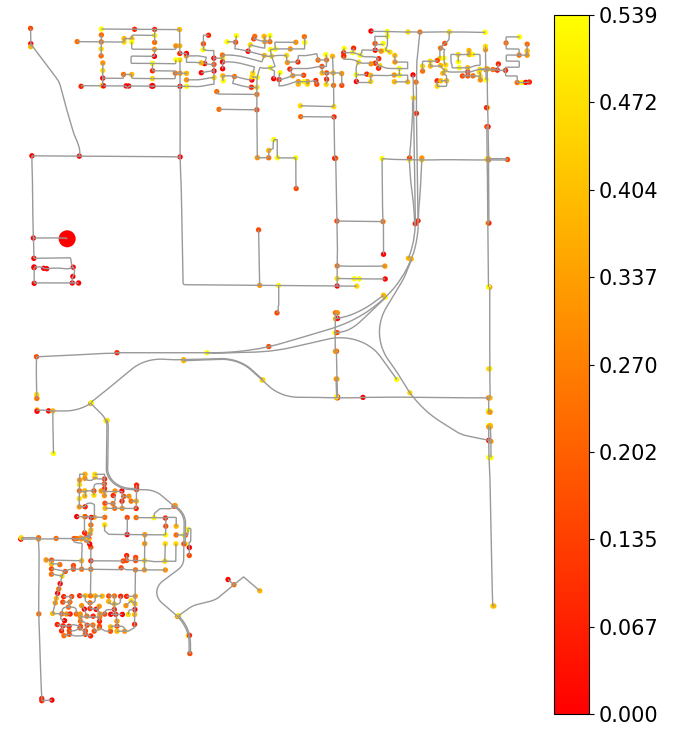}
        \caption{Distribution of real power (MW) among the distribution feeder nodes (population information excluded)}
        \label{fig:load_no_population}
    \end{figure}
\end{enumerate}


\subsection{Steady state voltage profile for sub-graph $g_{z,N}$}
\label{sec:power_flow}
\begin{table}[ht]
	\caption{List of Symbols}
	\begin{center}
		\begin{tabular}{|c|c|}
			\hline
			${\cal L}$ & Set of all lines for the graph $g_{z,N}$\\ \hline
			$p_i$, $q_i$ & Real and reactive power demand for node $i$ \\ \hline
			$r_{ij},x_{ij}$ & Resistance and reactance for line $(i,j)$\\ \hline
			$P_{ij},Q_{ij}$ & Real and reactive power flows from bus $i$ to $j$\\ \hline
			$v_i$ & Voltage magnitude for node $i$\\ \hline
		\end{tabular}
	\end{center}
	\label{table:symbols}
\end{table}
The framework has a radial graph $g_{z,N}$ of lines with known length, a substation node, and the real and reactive power demand at each node. Bus $0$ denotes the substation node or the point of common coupling (PCC), with a predefined reference voltage ($v_0$). Additional parameters including line resistance and reactance are required to run a steady state power flow.

Following the lossless linearized distribution power flow or \DistFlow equations from \cite{farivar2013equilibrium} and using symbols introduced in Table \ref{table:symbols}, for every  $(i,j) \in \cal L$ of the radial distribution feeder:
\begin{subequations} 
\label{eq:original_equations}
\begin{align}
P_{ij}&=p_{j}+\sum_{k:(j,k)\in{\cal L}} P_{jk} \label{eq:pdistflow} \\
Q_{ij}&=q_{j}+\sum_{k:(j,k)\in{\cal L}} Q_{jk} \label{eq:qdistflow} \\
v_j^2-v_i^2&=-2(r_{ij}P_{ij}+x_{ij}Q_{ij})\label{eq:vj2-vi2}
\end{align}
\end{subequations}

\begin{algorithm}[]
    \caption{Algorithm for choosing appropriate cable with minimum number of parallel cables }\label{alg:parallel_cable}
        \begin{algorithmic}[1]
        \Statex \textbf{Input~~:} $P_{i,j},~Q_{i,j}$ for all lines, cable database ($\cal C$), maximum number of parallel allowed$= l_{\max}$
        \Statex \textbf{Output~~:} Resistance and reactance for each line of the graph ($g_{z,N}$)
        \State Sort the cable database in ascending order of MVA capacity ($S_{c}$) of the cables
        \State $\bm{c} \leftarrow $ vector containing the $S_c$ values for all cables
        \State $\bm{l} \leftarrow 1\dots l_{\max}$
        \State $\bm{C_s} \leftarrow  \bm{c}~\bm{l}^T$
        \For {each line segment $(i,j)\in \cal L$}
            \State calculate MVA flow $(S_{i,j})$ through line $i,j$ using $P_{i,j}$ and $Q_{i,j}$
            \State flag $\leftarrow$ False
            \For {each column in $\bm{C_s}$}
                \For {each row in $\bm{C_s}$}
                    \If{$\bm{C_s}[row][column] > S_{i,j}$}
                        \State calculate $r_{i,j}$ and $x_{i,j}$ using the resistance and reactance of the cable and the length of line $i,j$
                    \State flag $\leftarrow$ True
                    \Break
                    \EndIf    
                \EndFor
                \If{flag}
                    \Break
                \EndIf
            \EndFor
        \EndFor
        \end{algorithmic}
    \end{algorithm}

Equation \eqref{eq:pdistflow}-\eqref{eq:qdistflow} can be used to calculate the vector of real and reactive power flows for every line using the nodal demand values. By using a predefined cable database (that includes impedance and apparent power (MVA) carrying capacity information) and the calculated line flows, the framework then follows Algorithm \ref{alg:parallel_cable} to find the appropriate cable that can meet the power flow through a line with minimum number of parallels. Assigning cables allows the calculation of $r_{ij}$ and $x_{ij}$ for all lines, and  equation \eqref{eq:vj2-vi2} is then used to calculate nodal voltages using $v_0 = 1.00.$

If the minimum of the node voltages is less than a predefined voltage threshold ($v_{th}$), nodes with degree 1 are removed from $g_{z,N}$ and procedures described in \ref{assign_load} and \ref{sec:power_flow} are repeated for the modified version of the graph $g_{z,N}$. 


\subsection{Converting to \OpenDSS model}
\begin{figure}
    \centering
    \includegraphics[width=1\textwidth]{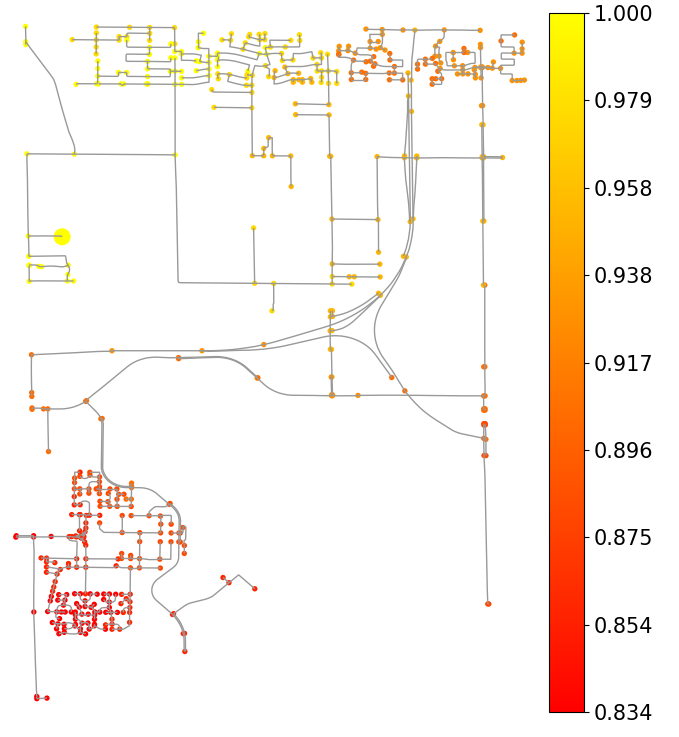}
    \caption{Steady state voltage profile using \OpenDSS}
    \label{fig:voltage_opendss}
\end{figure}

Output of the previous step is then converted to an \OpenDSS model to calculate a non-linear power flow solution for the positive sequence network model. The framework models all the loads as constant power loads with load and cable information retrieved from the graph $g_{z,N}$. Using the non-linear power flow provides the loss values across the network. The steady state voltage profile for the \OpenDSS model is shown in Fig. \ref{fig:voltage_opendss}.  

The \OpenDSS simulation returns loss values of $3.95832~+ j7.63604~\textnormal{MVA}$; hence active losses are $8.9\%$ of the load demand of $40.47~ + j 11.14~\textnormal{MVA}$. The non-linear power flow solution from \OpenDSS, which accounts for the losses, has larger currents, which are responsible for dragging the voltage profile lower than that calculated by the \DistFlow equation \eqref{eq:vj2-vi2}.  


\section{Extension to Multi-Phase Model}
The framework presented can be extended to multiple phases of an unbalanced distribution network by assuming the total load per phase is balanced across the entire distribution network. The following steps describe the procedure to include single-phase laterals to the developed distribution network.

\begin{enumerate}
    \item Select a total number of three-phase lines as a percentage of the total number of lines.
    \item Incrementally step through the power lines from the source to end-loads. Lines with more power flow are specified as three-phase lines until the total number of three-phase lines is reached. All other lines are specified as single-phase.
    \item Sum the real power demand across individual single-phase lateral to create the set of single phase nodes. 
    \item  Assign phases to the single-phase nodes by solving the optimization problem described in \ref{eq:optimization} to achieve balance among total real power load value per phase.
\end{enumerate}
\subsection{Nomenclature}
\begin{itemize}
    \item $\mathcal{C}_{N}$ represents the set of single phase nodes.
    \item $d_{AB}, d_{BC}, d_{CA}$ represent the difference of total power of phase A and phase B; phase B and phase C ; phase C and phase A respectively. 
    \item $u_{n}^{A}, u_{n}^{B}, u_{c}^{C}$ represent the binary variables corresponding to individual phase for node $n$.
    \item $p_n$ represents the real power demand of node $n \in \mathcal{C}_{N}$.
\end{itemize}
 
\subsection{Optimization Formulation}
\label{eq:optimization}
\begin{subequations}
\begin{align}
    &\min_{d_{AB},d_{BC},d_{CA},u_{n}^{A},u_{n}^{B},u_{n}^{C}}~~d_{AB} + d_{BC} + d_{CA}  \\
    &\textnormal{subject to} \nonumber \\
    &u_{n}^{A}, u_{n}^{B}, u_{c}^{C} \in \{0,1\}\\
    &u_{n}^{A} + u_{n}^{B} + u_{n}^{C} = 1 ~~\forall n \in \mathcal{C}_{N} \\
    &\sum_{\forall n \in \mathcal{C}_{N}} u_{n}^{A}~p_n = P^A  \\
    &\sum_{\forall n \in \mathcal{C}_{N}} u_{n}^{B}~p_n = P^B \\ 
    &\sum_{\forall n \in \mathcal{C}_{N}} u_{n}^{C}~p_n = P^C \\
    &d_{AB} \geq 0 ~;~ d_{BC} \geq 0 ~;~ d_{CA} \geq 0 \\
    &-d_{AB} \leq P^A - P^B \leq d_{AB} \\
    &-d_{BC} \leq P^B - P^C \leq d_{BC} \\
    &-d_{CA} \leq P^C - P^A \leq d_{CA}
\end{align}
\end{subequations}

\begin{figure}
    \centering
    \includegraphics[scale=0.95]{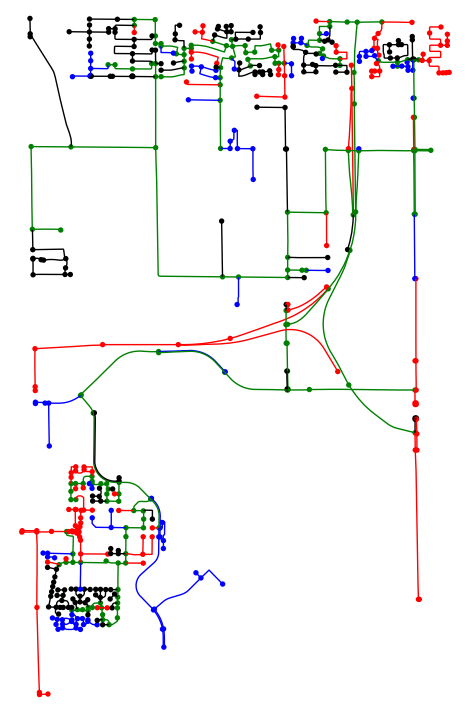}
    \caption{Introducing single-phase laterals with three-phase lines (green, black, red, and blue represent three-phase, A phase, B phase, C phase edges respectively)}
    \label{fig:three_single}
\end{figure}
Figure. \ref{fig:three_single} shows the distribution feeder with three-phase and single-phase edges.

\section{Conclusion and Future Work}
This paper introduces a framework to generate synthetic distribution feeders with real geo-spatial topologies. The methods employ a combination of street map data, US population census information, and prior work for synthetic transmission systems to reduce the burden of providing extensive inputs for the feeder generation. The software and data are public and freely available, allowing power systems researchers to develop thousands of realistic use cases. The framework used substations from literature on synthetic transmission systems to permit researchers to develop combined dataset including transmission and distribution systems. That joint work will allow researchers greater opportunity to run co-simulations of transmission and distribution systems, which is a growing area of research with few public dataset available to advance knowledge generation. 

This fundamental work is planned for expansion in a following study using unbalanced three-phase power flow, transformers and multiple voltage levels, and voltage supporting devices to more accurately depict a real distribution network. Voltages outside of ANSI standards will then be rectified. Such a model can also be exercised using time series information for the demand nodes and locations of distributed energy resources to generate many realistic use cases for how localized generation and storage can affect power flow and voltages in a distribution network.

\bibliographystyle{IEEEtran}  
\bibliography{bibfile,distflow} 

\end{document}

%% file: packages.tex
\usepackage{setspace}
\usepackage{xcolor}
\usepackage[flushleft]{threeparttable}
\usepackage{floatrow}
\usepackage{url}

\usepackage{graphicx}
\usepackage{amssymb}
\usepackage{epstopdf}
\usepackage{bm}
\usepackage{tcolorbox}
\usepackage{xcolor}
\usepackage{physics}
\usepackage{algpseudocode}
\usepackage{algorithm}
\usepackage{cite}
\usepackage{enumitem}
\usepackage{url}
\usepackage{amsfonts,amssymb,amsmath}
\usepackage{bm,theorem,color}
\usepackage{array}
\usepackage{calc}
\usepackage{cases} 
\usepackage{multirow}
\usepackage{tcolorbox}
\usepackage{epstopdf}
\usepackage{tikz}
\usepackage[european]{circuitikz}
\usetikzlibrary{shapes.multipart, positioning,calc,matrix}
\usetikzlibrary{patterns,arrows,decorations.pathreplacing}
\usetikzlibrary{shapes,arrows}
\usepackage{tabularx}
\usepackage{tabulary}
\usepackage{amsmath}
\usepackage{multirow}
\usepackage{bm}
\usepackage{commath}
\usepackage{cancel}	
\usepackage[font=footnotesize]{caption}
\usepackage{interval}
\usepackage{algorithm}
\usepackage{algpseudocode}
\usepackage{xcolor}
\definecolor{darkgreen}{RGB}{0,100,0}

\usepackage{mathtools}
\usepackage{newtxmath}
\usepackage{physics}

\usetikzlibrary{backgrounds}
\usepackage{pgfplots}
\pgfplotsset{compat=1.11}
\usetikzlibrary{decorations.pathmorphing,patterns}

\tikzstyle{startstop} = [rectangle, rounded corners, minimum width=3cm, minimum height=1cm,text centered, draw=black, fill=blue!30]
\tikzstyle{i} = [trapezium, trapezium left angle=70, trapezium right angle=110, minimum width=3cm, minimum height=1cm, text centered, text width=7cm, draw=black, fill=red!30]

\tikzstyle{o} = [trapezium, trapezium left angle=70, trapezium right angle=110, minimum width=3cm, minimum height=1cm, text centered, text width=7cm, draw=black, fill=green!30]

\tikzstyle{process} = [rectangle, minimum width=3cm, minimum height=1cm,text centered, text width=8cm ,  draw=black, fill=orange!30]
\tikzstyle{decision} = [diamond, minimum width=3cm, minimum height=0.5cm, text centered, inner sep=1pt, draw=black, fill=green!30]

\tikzstyle{joint} = [draw=black,circle,node distance=3cm,fill=yellow!20]

\tikzstyle{arrow} = [ultra thick,->,>=stealth]

\tikzset{
	connector/.style = {draw,circle,minimum width=1cm, minimum height=1cm, text centered,fill=yellow!20}
}

\pgfdeclarelayer{background}
\pgfdeclarelayer{foreground}
\pgfsetlayers{background,main,foreground}
\newcommand{\Break}{\State \textbf{break} }

\newcommand{\DistFlow}{DistFlow\ }

%% file: Figs/ProcessFlowv2.tex
\begin{tikzpicture}[node distance=2cm]
    \node (start) [startstop] {Start};
    \node (input) [i, below of=start, text width=4cm] {Substation location, MVA demand};
    \node (stepa) [process,below of=input] {\textbf{Step A} : Retrieve zip code ($\bm{Z}$) information};
    \node (stepb) [process,below of=stepa] {\textbf{Step B} : Generate the distribution feeder for $\bm{Z}$};
    \node (stepc) [process,below of=stepb,text width=8cm] {\textbf{Step C} : Distribute the load MVA within feeder nodes following a statistical distribution};
    \node (stepd) [process,below of=stepc] {\textbf{Step D} : Update distribution network with  appropriate cable \& run power flow using linearized distribution power flow equations};
    \node (stepe) [process,below of=stepd] {\textbf{Step E} : Convert to OpenDSS model \& run non-linear AC power flow};
    \node (stop) [startstop,below of=stepe] {Stop};
    
    \draw [arrow] (start) -- (input);
    \draw [arrow] (input) -- (stepa);
    \draw [arrow] (stepa) -- (stepb);
    \draw [arrow] (stepb) -- (stepc);
    \draw [arrow] (stepc) -- (stepd);
    \draw [arrow] (stepd) -- (stepe);
    \draw [arrow] (stepe) -- (stop);
\end{tikzpicture}